# SMILK, trait d'union entre langue naturelle et données sur le web

**Cédric Lopez[1,3], Molka Tounsi Dhouib[2], Elena Cabrio[2], Catherine Faron Zucker[2], Fabien Gandon[2], Frédérique Segond[1,4]**

1. *Viseo Technologies R&D*
   *4 avenue doyen Louis Weil, 38000 Grenoble, France*

2. *Université Côte d'Azur, Inria, CNRS, I3S*
   *Sophia Antipolis, France*
   *prenom.nom@inria.fr*

3. *Emvista – Montpellier, France*
   *prenom.nom@emvista.com*

4. *INALCO, ERTIM*
   *Paris, France*
   *prenom.nom@inalco.fr*

RÉSUMÉ. *Le laboratoire commun SMILK avait pour double sujet d'étude l'utilisation du traitement automatique du langage naturel pour aider à la construction et au liage de données sur le web et, à l'inverse, l'utilisation de ces données liées du web sémantique pour aider à l'analyse des textes et venir en appui de l'extraction de connaissances et l'annotation de contenus textuels. L'évaluation de nos travaux s'est focalisée sur la recherche d'informations portant sur des marques, plus particulièrement dans le domaine de la cosmétique. Cet article décrit chaque étape de notre approche : la conception de ProVoc, une ontologie pour décrire les produits et marques ; le peuplement automatique d'une base de connaissances reposant notamment sur ProVoc à partir de ressources textuelles hétérogènes; et l'évaluation d'une application prenant la forme d'un plugin de navigateur proposant des connaissances supplémentaires aux utilisateurs naviguant sur le web.*

ABSTRACT. *As part of the SMILK Joint Lab, we studied the use of Natural Language Processing to: (1) enrich knowledge bases and link data on the web, and conversely (2) use this linked data to contribute to the improvement of text analysis and the annotation of textual content, and to support knowledge extraction. The evaluation focused on brand-related information retrieval in the field of cosmetics. This article describes each step of our approach: the creation of ProVoc, an ontology to describe products and brands; the automatic population of a knowledge base mainly based on ProVoc from heterogeneous textual resources; and the evaluation of an application which that takes the form of a browser plugin providing additional knowledge to users browsing the web.*

MOTS-CLÉS : *web de données, ontologies, traitement automatique de la langue, données liées.*

## 1. Introduction

Cet article présente le fruit des travaux réalisés dans le cadre d'un laboratoire de recherche commun, le LabCom SMILK, entre l'équipe WIMMICS d'Inria et le centre de recherche de l'entreprise Viseo Technologies. L'ouverture et la mise à disposition de grands volumes de données publiques (*Open Data*) et l'application des principes du web à la mise en réseau des jeux de données en liant les données entre elles (*Linked Data*), ont créé de nouvelles opportunités et de nouvelles ressources pour les recherches en sciences du numérique. De nouveaux verrous scientifiques sont alors apparus, du fait de l'hétérogénéité des données et de leurs possibles mises en relation. Dans un tel contexte, l'objectif du LabCom SMILK a été de tirer le meilleur parti des technologies du traitement automatique du langage naturel (TALN) et du web des données ouvertes (*Linked Open Data*) pour : d'une part, extraire, analyser, lier et raisonner sur les données issues des ressources textuelles du web, en particulier des réseaux sociaux ; d'autre part, utiliser les données ouvertes du web en prenant en compte les structures et les interactions sociales afin d'améliorer l'analyse et la compréhension des ressources textuelles. En d'autres termes, il s'agit d'utiliser le TALN pour aider à la construction et au liage dans le web des données et d'utiliser le web des données pour aider à l'analyse des textes et venir en appui des technologies du TALN. D'un point de vue plus applicatif le LabCom SMILK visait à permettre la collecte précise, non ambiguë et non redondante d'informations sur le web, afin de faire émerger des liens qui n'apparaissent pas au premier abord.

Être en mesure de répondre à de telles ambitions nécessite de travailler selon les trois axes suivants :

– analyser des données pour en extraire du sens,

– établir des liens entre les données,

– enrichir les informations avec des données issues du web, en prenant soin de ne pas créer d'ambiguïté ou de redondance.

Nous avons choisi, dans le cadre de SMILK, de travailler à un scénario focalisé sur la recherche d'informations autour des marques, plus particulièrement dans le domaine de la cosmétique/beauté (domaine de compétence de plusieurs clients et potentiels fournisseurs de données de l'entreprise Viseo Technologies). Pour réaliser ce scénario, nous avons été amené à 1) développer et intégrer des approches du TALN pour extraire des informations, reconnaître et désambiguïser les entités nommées (ici essentiellement les marques de cosmétiques et les produits de ce domaine) présentes dans des textes traitant de cosmétique, 2) développer des approches du web sémantique avec notamment des algorithmes permettant de relier ces entités à une base de connaissances (ici DBpedia), et enfin 3) d'analyser le contenu textuel des réseaux sociaux, afin de visualiser toutes les informations recueillies. Le LabCom SMILK a donné lieu à quatre principaux résultats :

– un prototype de liage d'entités (*Entity Linking*) (Nooralahzadeh *et al.*, 2016),

- l'ontologie ProVoc qui exprime la connaissance liée aux produits (Lopez *et al.*, 2016), http://ns.inria.fr/provoc/
- les modules d'extraction d'information (reconnaissance d'entités nommées, extraction de relations entre ces entités) (Tounsi *et al.*, 2017),
- le plugin SMILK qui permet à l'utilisateur d'enrichir le contenu des pages web au cours de sa navigation sur le web (Lopez *et al.*, 2016).

Pour désambiguïser les entités nommées et les lier aux entrées de DBpedia, nous avons développé un algorithme non supervisé qui exploite d'une part des caractéristiques désormais largement adoptées par la communauté (distances entre chaînes de caractères, TF-IDF), et d'autre part, des caractéristiques basées sur la connectivité des entités dans DBpedia. Une des difficultés de ce scénario est que de nombreux produits n'ont pas d'entrée dans cette base de connaissances. Pour y remédier nous avons dû développer l'ontologie ProVoc (*Product Vocabulary*), une extension des ontologies GoodRelations (Hepp, 2008) et Schema.org, qui définit des propriétés spécifiques aux produits (par exemple, la composition des produits, l'impact qu'ils peuvent avoir sur la santé ou encore le packaging) et aux entités relatives (marques, gammes de produits, entreprises...). Le vocabulaire ainsi développé a permis de représenter les informations extraites automatiquement dans les pages web parcourues par l'utilisateur avec le plugin SMILK et de sauvegarder ces informations au fur et à mesure dans la base de connaissances. D'une façon très générale, lorsque l'utilisateur consulte une page web, le plugin SMILK détecte les entités nommées du domaine de la cosmétique et les surligne de couleurs différentes en fonction de leur type (noms de produits, gammes, marques, groupe de cosmétique, division d'un groupe). Chaque entité nommée est désambiguïsée et liée à la ressource DBpedia correspondante pour en extraire les données. Lorsque l'utilisateur clique sur une entité surlignée, un graphe est construit à la volée, permettant de visualiser les liens entre toutes les informations qui ont pu être extraites du texte, enrichies d'informations provenant de DBpedia et des réseaux sociaux (analyse d'opinions, nuages de mots...).

Dans ce qui suit, nous nous focalisons dans un premier temps sur la description du vocabulaire ProVoc (*cf.* section 2). Puis, nous présentons une des approches développées dans le cadre de SMILK pour extraire les relations entre entités (*cf.* section 3) : de sa construction à son utilisation dans le domaine de la cosmétique à travers le Plugin SMILK (*cf.* section 4). Enfin nous décrivons la tâche d'extraction de relations et l'application au sein plugin Chrome.

Pour la description de l'algorithme de liage des entités développé dans le cadre de SMILK, nous renvoyons le lecteur à (Nooralahzadeh *et al.*, 2016).

**2. ProVoc, une ontologie pour décrire les produits sur le web**

Le secteur du luxe, et en particulier le secteur de la cosmétique est historiquement clé pour l'entreprise Viseo avec de nombreux clients comme L'Oréal, LVMH, Séphora, ou L'Occitane. En conséquence, les scénarios d'application de SMILK ont été définis autour du concept de produit dans le domaine de la cosmétique (*cf.* section 4). Nous avons été attentifs à la construction d'une ontologie qui réponde à ce besoin applicatif tout en étant néanmoins la plus générique possible, et en particulier non exclusivement dédiée à la cosmétique (Lopez *et al.*, 2016).

*2.1. Contexte*

Cette dernière décennie a vu le nombre de produits disponibles dans le commerce largement augmenter. Par exemple, le nombre de références pour l'alimentation infantile a augmenté de 58 %, le nombre de références pour le café torréfié a augmenté de 81 %, quant aux produits de beauté ils ont connu une augmentation de 42 %[1].

Devant une telle masse de produits, le client aborde le problème de décision d'achat selon ses propres critères : le prix, la marque, la composition du produit, la provenance, la présentation, la qualité, l'appréciation globale par la communauté, les comparatifs, les avis de ses proches, etc. Pour aider à la prise de décision, de nouvelles applications ont vu le jour, notamment dans le domaine de l'alimentation, telles que ShopWise qui se concentre sur la composition de plus de 25 000 produits alimentaires, EcoCompare qui permet d'évaluer des produits en fonction des critères d'éco-responsabilité (environnement, sociétal, santé), ou SkinDeep qui recense les ingrédients potentiellement dangereux dans les produits cosmétiques.

Alors que les consommateurs cherchent de plus en plus à acquérir des informations sur des produits, les ontologies ouvertes disponibles au format du web sémantique proposent une représentation pertinente dans le contexte du e-commerce mais la couverture de la représentation d'informations relatives aux produits eux-mêmes reste faible.

Depuis 2009, Google permet l'enrichissement des résultats de son moteur de recherche en proposant notamment les *rich snippets* dédiés aux e-commerçants. Les *rich snippets* ont pour objectif de fournir une indexation et un affichage plus pertinent pour les sites web dans les résultats du moteur (images, caractéristiques, évaluations, localisation, *etc.*) et permettent aux développeurs d'améliorer le trafic et le référencement de leurs pages. Cependant les *rich snippet* et les ontologies associées ne couvrent pas nos besoins : le vocabulaire utilisé est un sous-ensemble du vocabulaire de schema.org (précisément des classes Product, Offer, AggregateOffer) lui-même inspiré de GoodRelations (Hepp, 2008) qui concerne essentiellement des scénarios de e-Commerce.

En effet, GoodRelations, l'ontologie au format du web sémantique la plus utilisée dans le monde du e-Commerce (Ashraf *et al.*, 2011), se positionne comme le vocabulaire le plus puissant qui permette de publier des détails sur les produits et services. Elle est fondée sur la structure agent-objet-promesse-compensation (agent :

---

1. Ces pourcentages sont issus de http://www.journaldunet.com/

personne ou organisation ; promesse : transfert de la propriété d'un objet par exemple ; objet : un objet ou un service ; compensation : par exemple un montant monétaire). Cette ontologie est donc orientée vers les transactions en ligne plus que vers l'aide à la décision d'achat. Adoptant un angle de vue différent, mais compatible, nous proposons l'ontologie ProVoc (*Product Vocabulary*), développée dans le cadre du projet SMILK.

Contrairement à GoodRelations qui se place dans le contexte du e-Commerce, Provoc s'intéresse en général à la description et l'organisation de catalogues de produits. De façon complémentaire à GoodRelations, ProVoc présente deux intérêts :

1) une représentation plus fine des produits qui permet de répondre à des requêtes telles que "Quelles gammes contiennent les produits que je recherche ?",

2) la possibilité de tisser des liens vers des informations *a priori* hors catalogue. On peut, par exemple, établir des liens vers des informations relatives à la santé *via* la composition des produits ("Quels aliments contiennent des ingrédients néfastes pour la santé ?"), ou établir des liens vers des ontologies telles que FOAF (Brickley, Miller, 2010) et des bases de connaissances telles que DBpedia. On ouvre ainsi le champ des requêtes possibles : "Quels sont les parfums représentés par des actrices qui ont joué dans Star Wars III ?" et par le biais de GoodRelations "Qui les vend et sous quelles modalités ?". L'objectif de ProVoc est donc de représenter, publier et relier des informations issues de catalogues de produits à d'autres données ouvertes et liées sur le web ou internes à un web sémantique d'entreprise. Dans la section suivante, nous adoptons la méthodologie de Uschold, Gruninger (1996) pour construire notre ontologie à partir de scénarios issus de clients de la société Viseo. Ces scénarios mettent en avant des situations impossibles à représenter avec GoodRelations que nous résolvons avec ProVoc. La section 2.3 donne un aperçu des entités et relations de ProVoc et nous discutons leur positionnement vis-à-vis de GoodRelations. Les choix de langage et l'évaluation sont abordés dans la section 2.4.

*2.2. Modélisation et engagement ontologique*

Dans cette section, nous identifions les scénarios motivants et les questions de compétences en adoptant la méthodologie de (Uschold, Gruninger, 1996). Premièrement, nous identifions des scénarios issus de cas d'utilisation réels. À partir de ces scénarios, nous identifions les questions de compétences, c'est-à-dire les questions auxquelles notre ontologie doit être en mesure de répondre. Cette section fixe donc ce qu'il est convenu d'appeler notre engagement ontologique au sens de (Bachimont, 2000).

*2.2.1. Scénarios*

Les scénarios présentés ici sont issus des clients de Viseo du secteur de la cosmétique (L'Oréal, L'Occitane, et Moët Hennessy Louis Vuitton). Certains scénarios sont également issus de notre collaboration avec Beaute-test.com (Lopez

*et al.*, 2014), un guide d'achat des cosmétiques en ligne qui fournit près de 50 000 fiches produits et qui a pour objectif d'informer et de conseiller les internautes sur les produits de beauté. De tels scénarios ne peuvent pas être traités par les ontologies existantes. Ces scénarios motivent et délimitent la conception et la publication du vocabulaire ProVoc et permettent d'identifier des applications associées. Les scénarios 1 à 5 nécessitent un catalogue "précis" des produits ; le scénario 6 montre la nécessité d'établir des liens avec des données "hors catalogue".

– Scénario 1 : l'utilisateur recherche des informations sur les différents maillons d'une chaîne de distribution d'un produit. L'ontologie doit donc permettre de représenter les entités en mesure de fournir des produits à d'autres entités. Ces entités peuvent être une maison de fabrication, un distributeur, *etc.*, toute entité faisant partie intégrante des canaux de distribution.

– Scénario 2 : l'utilisateur cherche à se renseigner sur les composants d'un produit, par exemple, à savoir si ces composants sont néfastes pour la santé. Ces composants peuvent être chimiques, naturels, ou matériels, par exemple. L'ontologie doit donc permettre de représenter la composition des produits et leurs impacts sur la santé.

– Scénario 3 : l'utilisateur souhaite effectuer des recherches par gammes de produits attachées à une marque donnée. Il souhaite notamment pouvoir identifier et naviguer à l'intérieur des gammes. L'ontologie devra permettre de représenter les gammes de produits.

– Scénario 4 : l'utilisateur souhaite connaître les coffrets de produits qui contiennent un ou plusieurs produits en particulier. L'ontologie doit permettre de représenter un ensemble de produits vendus comme une unité. Par exemple un coffret de cosmétique ou un panier alimentaire.

– Scénario 5 : l'utilisateur souhaite connaître la cible d'un produit ou d'une marque (par exemple « animaux », « végétaux », « hommes », « femmes », « adultes » ou « enfants »). L'ontologie devra donc représenter la cible des produits et des marques.

– Scénario 6 : l'utilisateur souhaite identifier un produit représenté par une personne : actrice d'un film donné, mannequin, ou autres personnalités. L'ontologie devra donc représenter les personnes qui sont impliquées dans la publicité des produits.

### 2.2.2. *Questions de compétences*

Dresser une liste de questions de compétences est un moyen de déterminer les spécifications de l'ontologie (Grüninger, Fox, 1995). Une liste non exhaustive des questions de compétence issues des questions posées par les utilisateurs du forum Beauté-test qui a l'avantage d'être très actif et de couvrir de nombreux aspects de la cosmétique (produits de maquillage, produits de soins, parfums, *etc.*), est présentée ici :

- Q1 : Quels sont les fournisseurs d'un produit donné ? (Scénario 1)

- Q2 : Quel(le)s sont les produits | gammes | marques | divisions | sociétés qui présentent des risques pour la santé ? (Scénarios 1 et 2)

- Q3 : Quel(le)s sont les gammes | marques | divisions | sociétés qui ne commercialisent pas de produits contenant du propylène glycol ? (Scénarios 1 et 2)

- Q4 : Quelles sont les gammes de produits d'une marque donnée ? (Scénario 3)

- Q5 : Quel type de consommateur est ciblé par le/la produit | gamme | marque ? (Scénario 5)

- Q6 : Existe-t-il un coffret contenant le produit recherché ? A l'inverse, le produit inclus dans ce coffret est-il commercialisé unitairement ? (Scénario 4)

- Q7 : Quels sont les parfums représentés par des actrices ? (Scénario 6)

*2.2.3. Principales entités et relations de ProVoc*

Dans cette section nous présentons et discutons les principales entités et relations de ProVoc. Elles sont issues de la terminologie extraite des questions de compétences, lorsque celles-ci n'étaient pas représentées dans d'autres ontologies. Les primitives centrales du vocabulaire sont présentées en figure 1. Dans la suite, le préfixe « pv » désigne des ressources de ProVoc (espace de nommage http://ns.inria.fr/provoc/), et le préfixe « gr » désigne des ressources de GoodRelations (espace de nommage http://purl.org/goodrelations/v1# ).

L'espace de nommage et la publication de ProVoc respectent les principes des données liées sur le web et notamment la déréférenciation et la négociation de contenu par HTTP. Le vocabulaire ProVoc est référencé et intégré au catalogue LOV et le préfixe « pv » est enregistré sur prefix.cc.

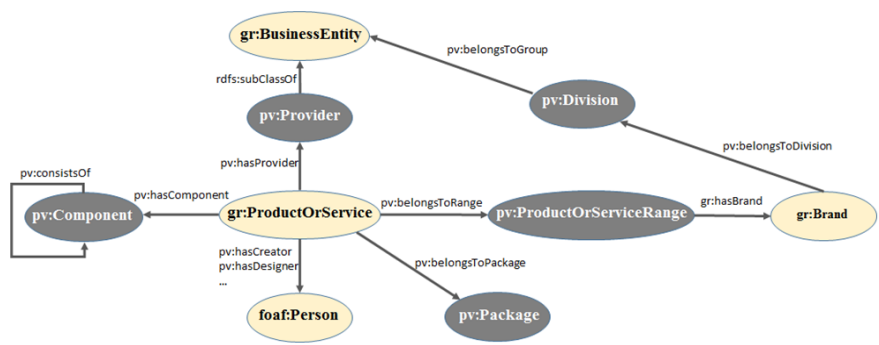

*Figure 1. Diagramme représentant les principales classes et propriétés de ProVoc (en gris et préfixées par « pv »)*

**Les principales entités et relations de ProVoc sont :**

**gr:isVariantOf** : GoodRelations définit des variantes de produits : « *A variant is a specialization of a product model and inherits all of its product properties, unless they are defined locally. This allows a very compact modeling of product models that vary only in a few properties.* » D'après GoodRelations, une variante d'un MacBook est par exemple un MacBook13Inch ou un MacBook15Inch qui varient par leur taille d'écran et la quantité de ports USB disponibles. Il s'agit d'héritage entre un produit "parent" et ses dérivés qui héritent des caractéristiques par défaut du produit "parent" à moins de redéfinir les valeurs localement, un peu à la manière d'une représentation orientée prototypes. Dans la version actuelle de GoodRelations, il existe une relation gr:isVariantOf qui doit nécessairement être utilisée entre deux modèles de produits ou services. Or, les gammes de produits peuvent difficilement être traitées comme un ensemble de dérivés d'un produit/modèle commun. Par exemple, Elsève est une gamme (de la marque L'Oréal Paris) proposant des shampooings avec des dérivés, la même gamme Elsève propose aussi des colorations avec des dérivés, des huiles avec des dérivés, *etc*. Ainsi, mis à part le trait commun qu'il s'agit de traitements pour les cheveux, ces produits ne partagent pas un prototype commun. L'utilisation de gr:isVariantOf entre certains produits proches, impliquerait que l'on obtienne plusieurs ensembles de produits apparentés, au détriment d'une gamme unique. *In fine,* les variantes de GoodRelations semblent pertinentes pour identifier des produits plus ou moins similaires, mais les gammes de produits ont d'après nous une toute autre vocation, notamment d'un point de vue fonctionnel et marketing, impliquant qu'elles doivent être définies par le fournisseur de façon non subjective. Or, gr:isVariantOf a une sémantique très large et subjective. Par exemple, rien n'empêche d'exprimer qu'une Renault Clio 4 est une variante d'une Ford Fiesta; pourtant elles ne sont pas de la même marque.

Pour ces raisons, nous introduisons dans ProVoc la notion de gammes de produits ou services pv:ProductOrServiceRange.

**pv:ProductOrServiceRange** : cette classe permet de représenter une gamme de produits de façon non subjective (fournie par l'expert) contrairement à gr:isVariantOf. Une instance de cette classe ne représente ni un produit, ni un modèle de produit. L'utilisation de cette classe permet d'affiner la représentation et le contenu des catalogues de produits et services tout en leur assurant un caractère objectif. On peut ainsi exprimer qu'un produit appartient à une gamme et que cette gamme est proposée par une marque. Par exemple que l'Huile extraordinaire est un produit de la gamme Elsève qui appartient à la marque L'Oréal :

```
@prefix pv: <http://ns.inria.fr/provoc#>
@prefix rdf: <http://www.w3.org/1999/02/22-rdf-syntax-ns#>
@prefix ex: <http://example.org#>

ex:Huile_extraordinaire_Universelle
        pv:belongsToProductOrServiceRange  ex:Elsève  .
ex:Elsève rdf:type pv:ProductOrServiceRange ;
        pv:belongsToBrand  ex:LOreal_Paris  .
```

**pv:Component** : une instance de cette classe représente un composant d'un produit. Un pv:Component peut être constitué d'autres pv:Component. Par exemple, les ingrédients d'un parfum.

**pv:Division** : une instance de cette classe représente une division (un sous-groupe) de BusinessEntity. En effet, une organisation est parfois divisée en plusieurs divisions, et chaque division propose des marques différentes. GoodRelations lie gr:BusinessEntity directement à gr:Brand. Par exemple, L'Oréal Grand Public est une division du groupe L'Oréal.

**pv:Package** : un package est un ensemble de produits et/ou services. Par exemple, un coffret de cosmétique qui contient des crèmes, un parfum et un rouge-à-lèvres. Un autre exemple associant un produit et un service pourrait être un package contenant un traitement et un abonnement chez une esthéticienne. Cette classe est utilisée pour représenter un ensemble de produits qui est vendu unitairement. Cet ensemble représenté par pv:Package peut contenir des variantes d'un modèle de produit (utilisation de gr:isVariantOf), des produits appartenant à une même gamme (utilisation de pv:ProductOrServiceRange), ou des produits similaires (utilisation de gr:isSimilarTo). Par exemple :

```
@prefix gr: <http://purl.org/goodrelations/v1#>
@prefix rdf: <http://www.w3.org/1999/02/22-rdf-syntax-ns#>
@prefix ex: <http://example.org#>

ex:Degustabox1015
        rdf:type  pv:Package .
ex:MiniBiscuitsBanana
        pv:belongsToPackage  ex:Degustabox1015 ;
        pv:belongsToBrand  ex:Weetabix_Crispy_Minis ;
        gr:isSimilarTo  ex:MiniBiscuitsChocolate  ;
        gr:isVariantOf  ex:MiniBiscuitsModel  .
ex:MiniBiscuitsModel  rdf:type  gr:ProductOrServiceModel .
ex:CapuccinoCarambar  pv:belongsToPackage  ex:Degustabox1015 ;
        pv:belongsToBrand  ex:Maxwell_House  .
ex:ExtraitNaturelDeVanille
 pv:belongsToPackage  ex:Degustabox1015 ;
 pv:belongsToBrand  ex:Sainte-Lucie .
```

**pv:Provider :** le fournisseur est un type d'organisation. Il se distingue de l'organisation par le fait que les marques ne lui appartiennent pas; il ne fait que les commercialiser. Le fournisseur propose des marques à la vente qui n'appartiennent pas toujours à la même organisation. Exemple : Carrefour vend des produits des entreprises L'Oréal et Design Paris. D'autres ressources viennent enrichir ce modèle, notamment pour assurer les liaisons entre les données du catalogue et les données hors-catalogue. Par exemple, les classes pv:Ambassador, pv:Designer et pv:Model sont des sous-classes de foaf:Person qui représentent des fonctions qu'exercent des personnes impliquées dans la chaîne de production, la commercialisation, ou la communication du produit (par exemple les égéries des marques). L'introduction de ces classes dans ProVoc permet d'établir les liens entre le catalogue de produit et des données hors catalogue (ici en se référant à des personnes), provenant de bases de connaissances telles que DBpedia.

*2.3. Choix du langage et évaluation*

ProVoc a été éditée avec le logiciel Protégé. L'ontologie et la description de ses ressources et propriétés sont publiées selon les principes des données liées sur le web et le schéma est identifié par l'URI http déréférençable http://ns.inria.fr/provoc#. L'ontologie ProVoc utilise les mêmes primitives que GoodRelations : owl:Ontology, owl:Class, owl:versionInfo, owl:DatatypeProperty, rdfs:subClassOf, rdfs:subPropertyOf, rdfs:comment, rdfs:domain, rdfs:range, rdf:datatype, rdf:type. De cette façon, les annotations effectuées avec ProVoc et GoodRelations sont dans le même fragment d'expressivité et peuvent être interprétées par un raisonneur RDF(S) qui sait traiter les éléments mentionnés ci-dessus.

Nous montrons, par le biais de quelques exemples de requêtes SPARQL, que toutes nos questions de compétences et leurs réponses sont exprimables dans le vocabulaire de ProVoc lié au préfixe « pv ».

Pour la question de compétence Q1 « Quels sont les fournisseurs d'un produit donné ? » issue du scénario 1, un exemple de question concrète est :

**Exemple 1.** Quels sont les fournisseurs de La Vie est Belle ? Sa formulation en SPARQL est :

```
SELECT DISTINCT ?fournisseur
WHERE {
        ex:La_Vie_est_Belle rdf:type gr:ProductOrService ;
        pv:hasProvider ?fournisseur .
        ?fournisseur rdf:type pv:Provider .
}
```

Pour la question de compétence Q2 « Quel(le)s sont les produits | gammes | marques | divisions | sociétés qui présentent des risques pour la santé ? » issue des scénarios 1 et 2, des exemples de questions concrètes sont :

**Exemple 2.** Quels sont les types de produits qui contiennent du linalool ? Sa formulation en SPARQL est :

```
SELECT DISTINCT ?type
WHERE {
        ?product pv:contains <http://fr.dbpedia/page/Linalool>
        ?product rdf:type ?type .
}
```

**Exemple 2bis.** Quelles sont les gammes de produits de la marque L'Oréal Paris qui présentent des risques pour la santé ? Sa formulation en SPARQL est :

```
SELECT DISTINCT ?range
WHERE {
        ?range pv:belongsToBrand ex:LOreal_Paris .
        ?product pv:belongsToRange ?range .
        ?product pv:contains ?component .
        ?component pv:healthimpact ?healthimpact .
}
```

Pour la question de compétence Q3 « Quel(le)s sont les gammes | marques | divisions | sociétés qui ne commercialisent pas de produits contenant du propylène glycol ? » issue des scénarios 1 et 2, un exemple de question concrète est :

**Exemple 3.** Quelles sont les marques qui ne commercialisent pas de produits contenant du propylène glycol ? Sa formulation en SPARQL est :

```
SELECT DISTINCT ?brand
WHERE {
        ?brand rdf:type gr:Brand .
        FILTER NOT EXISTS {
        ?x  rdf:type gr:Brand
        ?range pv:hasBrand  ?x .
        ?product pv:belongsToRange ?range .
        ?product pv:hasComponent ex:Propylene_glycol . }
}
```

Pour la question de compétence Q4 « Quelles sont les gammes de produits d'une marque ? » issue du scénario 3, un exemple de question concrète est :

**Exemple 4.** Quelles sont les gammes de produits proposées par L'Oréal Paris ? Sa formulation en SPARQL est :

```
SELECT DISTINCT  ?range
WHERE {
        ?range pv:belongsToBrand  ex:LOreal_Paris
}
```

Pour la question de compétence Q5 « Quel type de consommateur est ciblé par le/la produit | gamme | marque ? » issue du scénario 5, un exemple de question concrète est :

**Exemple 5.** Quel type de consommateur est ciblé par la gamme Elsève ? Sa formulation en SPARQL est :

```
SELECT DISTINCT  ?consumer
WHERE {
        ex:Elseve pv:hasTarget ?consumer
}
```

Pour la question de compétence Q6 « Existe-t-il un coffret contenant le produit recherché ? A l'inverse, le produit inclus dans ce coffret est-il commercialisé unitairement ? » issue du scénario 4, un exemple de question concrète est :

**Exemple 6.** Quels sont les coffrets distribués par Sephora qui contiennent le parfum La vie est Belle ? Sa formulation en SPARQL est :

```
SELECT DISTINCT  ?package
WHERE {
        ex:La_Vie_est_Belle  pv:belongsToPackage  ?package .
        ?package  pv:hasProvider  ex:Sephora
}
```

Pour la question de compétence Q7 « Quels sont les parfums représentés par des actrices ? » issue du scénario 6, un exemple de question concrète est :

**Exemple 7.** Quels sont les produits représentés par des acteurs qui ont joué dans Star Wars III ? Sa formulation en SPARQL est (avec les préfixes dbo:http://dbpedia.org/ontology/ et dbp:http://dbpedia.org/page/) :

```
SELECT DISTINCT  ?product
WHERE {
        dbp:Star_Wars_Episode_III:_Revenge_of_the_Sith  dbo:starring  ?actor .
        ?product  pv:hasRepresentative  ?actor
}
```

En intégrant le vocabulaire de GoodRelations, ces requêtes peuvent être enrichies par des interrogations relatives au e-Commerce. Par exemple, en cosmétique, « Qui vend, et sous quelles conditions, des coffrets destinés aux hommes contenant des produits appartenant à des gammes qui n'utilisent aucun composant néfaste pour la santé ? ». Afin de répondre à ces questions, nous avons développé une approche d'extraction d'information dans le but de peupler une base de connaissances fondée principalement sur ces vocabulaires. La section suivante décrit ces travaux.

**3. Peuplement d'une base de connaissances avec ProVoc**

Un objectif de SMILK est d'étudier le potentiel des techniques du TALN à des fins de peuplement de bases de connaissances. Le vocabulaire ProVoc sert de support à cette expérience. Contrairement à une approche visant à concevoir une ontologie à partir des résultats obtenus à la phase d'extraction d'information (Kiryakov *et al.*, 2004), ou d'adapter l'ontologie aux informations à extraire (Alec *et al.*, 2014) notre ontologie a été conçue indépendamment de la phase d'extraction, dans la lignée de (Amardeilh *et al.*, 2005). Notre objectif est d'évaluer le niveau de difficulté d'une tâche d'extraction d'information, plus précisément d'extraction de relations, guidée par une ontologie construite sans *a priori* vis-à-vis de cette tâche.

Les expériences décrites dans la suite considèrent que le liage d'entités est déjà réalisé par un module décrit dans (Nooralahzadeh *et al*., 2016). Ce module a pour fonction de lier les mentions du texte avec les entités représentées dans les bases de connaissances en tenant compte à la fois du contexte local et du contexte global de chaque mention et entité. Deux mentions ambiguës dans le texte peuvent être désambiguïsées si elles sont représentées par des entités proches dans le graphe des connaissances. Dans l'exemple donné en figure 2, la mention « YMCA » est désambiguïsée grâce au contexte qui contient la mention « Martti Ahtisaari » (entre autres) : cette dernière est liée à l'unique entrée dbr:Martti_Ahtisaari dans la base de connaissances, ce qui privilégie les entités qui y sont liées, notamment dbr:YMCA.

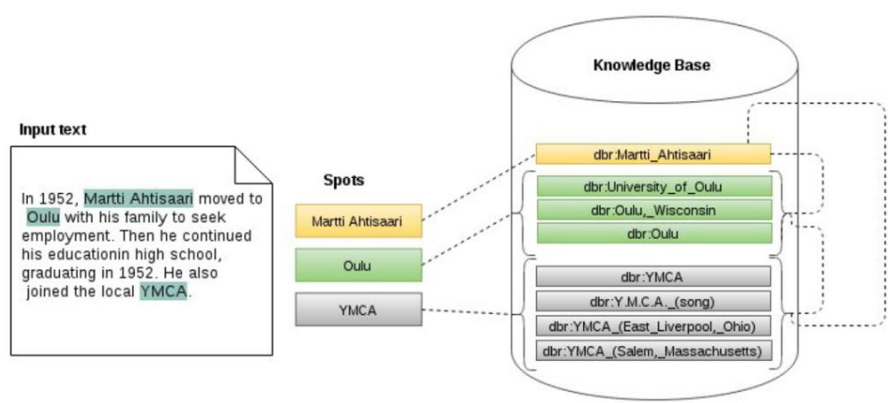

*Figure 2. Modèle de graphe pour le liage des données*

L'étape d'extraction de relations produit des triplets RDF qui sont insérés dans une base de connaissances après validation semi-automatique de ces triplets (*cf.* section 4). Nous nous focalisons sur l'étape d'extraction de relations.

Une première expérience a consisté à développer un module de peuplement automatique de la base de connaissances à partir de tweets « non bruités » fondée sur des règles linguistiques (Lopez *et al.*, 2017). L'intérêt de l'expérience résidait dans le fait que la taille limitée des tweets (140 caractères) a pour conséquence de décrire l'essentiel de l'information : les éléments textuels situés entre le terme représentant la relation (par exemple, un verbe ou sa forme nominalisée) et les entités à sa droite et à sa gauche sont donc réduits. D'une part, les règles d'extraction se basent sur une syntaxe « simple » et une quantité d'information limitée ; d'autre part, le contexte peut manquer lors de la désambiguïsation. L'évaluation met en évidence l'utilisation du conditionnel, de négation, de coréférence, d'humour et de démentis (fausses informations), qui sont des points difficiles à traiter mais qui permettent néanmoins d'obtenir des résultats exploitables en situation réelle grâce à une précision élevée (précision : 0,90 ; rappel : 0,68).

Une seconde expérience est celle décrite dans la suite de cette section. Son intérêt est d'expérimenter le peuplement automatique à partir de textes contenant plus de contexte que les tweets et qui correspondent à la situation réelle dans laquelle est immergé l'utilisateur parcourant les pages web dans le cadre de notre application. La désambiguïsation des entités est « facilitée » mais les règles d'extraction se basent sur une syntaxe plus complexe et une quantité d'information accrue. Plus précisément, dans ce travail, nous nous sommes focalisés sur l'extraction des trois relations pv:hasComponent, pv:hasReserntative et pv:hasFragranceCreator de l'ontologie ProVoc, et nous avons cherché à extraire ces relations à partir de deux types de corpus : (i) un corpus L'Oréal contenant des articles journalistiques et (ii) un corpus web issu d'une extraction manuelle de phrases issues du web par le biais du moteur de recherche Google.

### *3.1. Travaux antérieurs*

La tâche d'extraction des relations consiste à détecter des liens sémantiques qui existent entre différentes entités. Les différentes approches existantes peuvent être classées en trois catégories.

Une première approche, supervisée, consiste à considérer le problème de l'extraction de relations comme un problème de classification, et à utiliser un classifieur linéaire, auquel il faut fournir un ensemble d'exemples positifs et un ensemble d'exemples négatifs pour l'apprentissage. Par exemple, les méthodes dites *feature-based* et *kernel-based* décrites dans (Nebhi, 2013) adoptent cette approche.

Une deuxième approche, semi-supervisée, permet au système d'apprendre itérativement des patrons et des instances à partir d'un nombre réduit d'instances : les patrons sont utilisés pour extraire des nouvelles relations à partir de l'ensemble de données, qui sont ajoutées à l'ensemble des exemples. Cette opération est répétée jusqu'à ce qu'aucune nouvelle relation ne puisse être apprise à partir de l'ensemble de données. Les systèmes DIPRE, Snowbal, supervision distance, BOA décrits par Kumar et Manocha (2015) adoptent cette approche.

Une troisième approche, non supervisée, est la génération des patrons d'extraction telle que décrite dans (Nebhi, 2013). Cette méthode consiste à collecter des paires de mots avec les chaînes de caractères (*string*) qui les séparent, pour calculer la cooccurrence de termes ou pour générer les patrons. RdfLiveNews est un système décrit dans (Gerber *et al.*, 2013) qui adopte cette approche pour générer une base de connaissances RDF à partir de textes.

### *3.2. Approche*

Le point de départ de ce travail est (Tounsi *et al.*, 2017) qui consiste en une première expérimentation de l'extraction des relations dans le domaine de la cosmétique en utilisant uniquement des relations de dépendances syntaxiques. Nous présentons ici une deuxième expérimentation basée sur l'introduction du lexique au sein de ces règles syntaxiques. Pour cela, nous avons utilisé les mêmes outils que dans le travail préliminaire, notamment pour reconnaître les entités nommées et pour le repérage et l'étiquetage des dépendances syntaxiques entre les mots. Nous avons utilisé (i) Renco décrit par (Lopez *et al.*, 2014), basé sur des règles lexico-syntaxiques, qui permet d'extraire à partir des textes français des produits, leurs ingrédients, des gammes de produits, des marques, des divisions et des groupes, (ii) Holmes Semantic Solutions[2], basé sur une approche hybride (symbolique et statistique) pour reconnaître les noms de personnes, (iii) Talismane (Urieli, 2013) pour analyser les dépendances syntaxiques, (iv) le module de liage des entités avec la base de connaissances DBpedia développé dans le cadre du projet SMILK (Nooralahzadeh *et al.*, 2016).

Notre approche d'extraction se compose de deux modules : un premier module basé sur la définition de règles lexico-syntaxiques en injectant du lexique au sein des règles syntaxiques. Par exemple, nous avons défini les deux règles suivantes pour extraire l'objet et le sujet de la propriété pv:hasFragranceCreator de phrases telles que « Eau Mega composé par Olivier Polge, ce bouquet de jasmin et de pivoine cible les femmes qui voient la vie en grand. ». La règle pour extraire l'objet de la propriété est :

**SI** lemme (verbe)= composer | créer | imaginer | élaborer | travailler | retravailler et dep(V,P)=p_obj **ET** dep(P,(NC | NPP))=prep **ET** type(NC | NPP)=person **ALORS** (NC | NPP) est l'objet de la propriété pv:hasFragranceCreator.

La règle pour extraire le sujet de la propriété est :

**SI** lemme (verbe)= composer | créer | imaginer | élaborer | travailler | retravailler **ET** dep(V,(NC | NPP))=suj **ET** type (NC | NPP)=product **ALORS** (NC | NPP) est le sujet de la propriété pv:hasFragranceCreator.

Ce premier module a pour ambition d'améliorer la précision de règles purement syntaxiques, notamment par des règles telles que la suivante :

**SI** depRel(0,VPP)=mod **ET** depRel(VPP,NC)=(mod ou suj ou prep) **ET** depRel(NC,(P ou P+D))=dep **ET** depRel((P ou P+D),NPP)=prep **ET** type (NPP)=PER **ALORS** (NPP)=(FragranceCreator|| Representative).

Cette dernière règle, trop générale, pourrait en effet être indifféremment utilisée pour l'extraction des propriétés pv:hasFragranceCreator et pv:hasRepresentative, avec donc une faible précision.

Le deuxième module vient en complément du premier module pour traiter les phrases pour lesquelles les règles lexico-syntaxiques n'ont pas permis d'extraire de relations. L'approche repose sur :

– la définition d'un lexique associé à chaque relation : le lexique a été élaboré en analysant l'ensemble d'apprentissage. Particulièrement, il s'agissait de constituer l'ensemble des mots qui séparent le domaine et le co-domaine.

---

2. http://www.ho2s.com/fr/

– l'application d'un outil de reconnaissance d'entités nommées : les types de ces entités (personne, marque, produit, *etc.*) correspondent aux domaines et co-domaines de la relation que l'on cherche à extraire. Dans le cas où le domaine et le co-domaine ont été identifiés, les éléments textuels se situant entre la signature et le prédicat sont recherchés dans le lexique. Si la recherche s'avère fructueuse, la relation est extraite.

Finalement, nous avons écrit 11 règles pour la relation hasRepresentative, 7 règles pour la relation hasFragranceCreator et 23 règles pour la relation hasComponent. Par exemple, nous avons défini pour la relation pv:hasRepresentative le lexique « incarner, représenter, symboliser, ambassadeur, égérie, acteur, mannequin, star, icône, visage, image, ... ». Le système recherche d'abord dans une phrase une entité nommée de type pv:Product (le domaine de la propriété pv:hasRepresentative et une entité nommée de type foaf:Person (le co-domaine de la propriété), et en cas de succès, il recherche un élément du lexique défini. Par exemple, pour la phrase « Nina de Nina Ricci plonge dans un monde féerique avec le mannequin suédois Frida Gustavsson. » nous pouvons extraire la relation pv:hasRepresentative grâce à :

– la présence de l'entité nommée « Nina » de type pv:Product qui représente le sujet de la relation,

– l'entité nommée Frida Gustavsson de type foaf:Person qui représente l'objet et le mot « mannequin » qui se trouve dans le lexique associé à la relation pv:hasRepresentative.

### 3.3. *Expérimentations*

Nous avons testé notre approche sur deux corpus différents :

– le corpus L'Oréal est issu de l'outil Factiva[3] et contient des articles journalistiques contenant la mention L'Oréal. Ce corpus est composé de 392 phrases issues de différents journaux tels que Aufeminin.com, Cosmétique Hebdo et Cosmétique Mag. Ces phrases peuvent contenir ou non des relations ProVoc. Ce corpus contient 79 relations de type hasComponent, 38 relations de type hasFragranceCreator et 44 de type hasRepresentative ;

– le corpus web est issu d'une extraction manuelle de phrases issues du web par le biais du moteur de recherche Google. Nous avons cherché le nom d'un parfum dans Google, par la suite nous avons choisi les dix premiers liens après les annonces tels que des liens vers des sites commerciaux ou des blogs cosmétiques. Ce corpus est composé de 119 phrases. Il contient 69 relations de type pv:hasFragranceCreator, 62 relations de type pv:hasComponent, 81 relations de type pv:hasRepresentative.

Nous avons tout d'abord évalué notre système d'extraction de relations dans le but d'estimer l'impact de l'introduction du lexique au sein de règles syntaxiques. Nous comparons la précision et le rappel obtenus avec notre premier module selon les résultats décrits dans (Tounsi *et al.,* 2017) obtenus avec un ensemble de règles purement syntaxiques.

---

3. https://www.dowjones.com/products/factiva/

Sur le corpus du web, et pour la relation pv:hasFragranceCreator la valeur du rappel augmente de 0.43 à 0.45 et celle de la précision de 0.45 à 0.67. De même, le rappel de l'extraction de la relation pv:hasRepresentative a augmenté de 0.30 à 0.36 et la précision de 0.52 à 0.96. Nous avons également constaté une augmentation du rappel pour l'extraction de la relation pv:hasComponent de 0.20 à 0.30 et une diminution de 1 à 0.79 pour la précision.

Sur le corpus de L'Oréal la précision de notre système passe de 0.20 à 0.66 pour la relation pv:hasFragranceCreator et de 0.18 à 0.76 pour la relation pv:hasRepresentative. Par contre la valeur de la précision pour la relation pv:hasComponent reste à 0.25. La valeur du rappel pour la relation pv:hasComponent est diminuée de 0.06 à 0.02. Pour la relation pv:hasFragranceCreator la valeur du rappel est passée de 0.20 à 0.23 et de 0.18 à 0.24 pour la relation pv:hasRepresentative.

Dans un deuxième temps, nous avons évalué l'intérêt d'introduire notre deuxième module en complément du premier.

Nous avons obtenu une augmentation des valeurs du rappel pour le corpus du web : de 0.45 à 0.59 pour la relation pv:hasFragranceCreator, de 0.36 à 0.48 pour la relation pv:hasRepresentative, et de 0.3 à 0.5 pour la relation pv:hasComponent. Par contre, nous observons une diminution des valeurs de la précision : de 0.67 à 0.53 pour la relation pv:hasFragranceCreator, de 0.96 à 0.69 pour la relation pv:hasRepresentative et de 0.79 à 0.67 pour la relation pv:hasComponent.

Pour le corpus de L'Oréal, les valeurs du rappel ont augmenté pour toutes les relations : de 0.02 à 0.21 pour pv:hasComponent, de 0.23 à 0.32 pour pv:hasFragranceCreator et de 0.24 à 0.34 pour pv:hasRepresentative. Par contre les valeurs de la précision ont diminué: de 0.25 à 0.15 pour pv:hasComponent, de 0.66 à 0.32 pour pv:hasFragranceCreator et de 0.76 à 0.48 pour pv:hasRepresentative.

*Tableau 1. Récapitulatif des résultats sur le corpus du web*

| Propriétés | Méthodes d'extraction | Rappel | Précision |
|---|---|---|---|
| pv:hasComponent | Règles syntaxiques | 0.2 | 1 |
| pv:hasFragranceCreator | | 0.43 | 0.45 |
| pv:hasRepresentative | | 0.3 | 0.52 |
| pv:hasComponent | Règles lexico-syntaxiques | 0.3 | 0.79 |
| pv:hasFragranceCreator | | 0.45 | 0.67 |
| pv:hasRepresentative | | 0.36 | 0.96 |
| pv:hasComponent | Définition du lexique | 0.5 | 0.67 |
| pv:hasFragranceCreator | | 0.59 | 0.53 |
| pv:hasRepresentative | | 0.48 | 0.69 |

*Tableau 2. Récapitulatif des résultats sur le corpus de l'Oréal*

| Propriétés | Méthodes d'extraction | Rappel | Précision |
|---|---|---|---|
| pv:hasComponent | Règles syntaxiques | 0.06 | 0.25 |
| pv:hasFragranceCreator | | 0.21 | 0.20 |
| pv:hasRepresentative | | 0.11 | 0.18 |
| pv:hasComponent | Règles lexico-syntaxiques | 0.02 | 0.25 |
| pv:hasFragranceCreator | | 0.23 | 0.66 |
| pv:hasRepresentative | | 0.24 | 0.76 |
| pv:hasComponent | Définition du lexique | 0.21 | 0.15 |
| pv:hasFragranceCreator | | 0.32 | 0.32 |
| pv:hasRepresentative | | 0.34 | 0.48 |

Les tableaux 1 et 2 synthétisent les résultats des différentes méthodes utilisées pour l'extraction des relations, respectivement sur le corpus du web et sur le corpus de L'Oréal.

La baisse des valeurs de précision constatée dans le deuxième module de notre approche s'explique par les erreurs de détection automatique du type des entités nommées. Nous pouvons aussi supposer que l'annotation d'un plus grand corpus pour l'apprentissage permettrait de définir davantage de règles lexico-syntaxiques et des lexiques plus riches, et d'augmenter ainsi les performances de l'extraction de relations.

Les scores de précision permettent d'envisager un système de validation semi-automatique des triplets générés, afin de les insérer dans la base de connaissances. Ce module de validation constitue une brique de l'application (section 4).

Nous constatons également un meilleur résultat de l'application du lexique sur le corpus web par rapport au corpus de L'Oréal. Cela peut s'expliquer par la richesse des phrases des journaux par rapport au corpus du web qui est en général issu des blogs cosmétiques qui ont des structures syntaxiques et un lexique similaire.

L'application de cette approche permet le peuplement d'une base de connaissances dont les données sont validées par un expert, assurant ainsi leur cohérence. Dans la section suivante, nous présentons une application qui utilise ces données liées pour enrichir les connaissances des utilisateurs naviguant sur le web.

**4. Application**

Pour valider nos recherches et en démontrer le potentiel, nous avons développé un prototype qui prend la forme d'un plugin de navigateur ayant pour objectif d'enrichir les connaissances des utilisateurs naviguant sur le web. L'enrichissement se fait en deux temps : (1) les entités d'intérêts sont reconnues automatiquement et surlignées sur la page web en cours de consultation; (2) l'utilisateur peut cliquer sur l'une des entités pour afficher un graphe RDF, lui-même déployable (holophrastage) en cliquant sur les nœuds. De cette façon, l'utilisateur enrichit ses connaissances selon ses intérêts personnels.

Dans un contexte de veille stratégique, notre cas d'application se concentre, comme nous l'avons dit précédemment, sur le secteur de la cosmétique, bien représenté chez Viseo. Dans un premier temps, nous décrivons la chaîne de traitement (section 4.1) puis nous décrivons les scénarios d'utilisation du point de vue d'un utilisateur (section 4.2) et du point de vue d'un expert de domaine (section 4.3).

*4.1. Chaîne de traitement*

La chaîne de traitement est représentée en figure 3. Deux acteurs sont impliqués : un utilisateur et un expert de domaine. L'utilisateur peut réaliser trois actions :

– **Consultation de la page web.** L'utilisateur consulte une page web dont le contenu textuel est extrait. Ce contenu « brut » subit un prétraitement (sélection du contenu de balises pertinentes, par exemple, puis suppression des balises). Le texte obtenu est analysé par le module de reconnaissance d'entités nommées Renco (Lopez *et al.*, 2014). Renco est un système à base de règles linguistiques (Lopez *et al.*, 2014). Les règles développées sont de type lexico-syntaxique, construites à partir des résultats de l'analyse syntaxique fournie par la société Holmes Semantic Solution (Ho2S) et fondées sur les principes de (McDonald, 1996) et (Hearst, 1992) qui considèrent les contextes droit, gauche, et interne de l'entité pour la désambiguïser.

– **Clic sur l'une des entités repérées.** Les entités repérées sont liées aux bases DBpedia et NetSent[4] par notre système de liage des données développé à partir du *framework* Dexter (Ceccarelli *et al.*, 2013), pour lequel nous avons retenu notre approche de propagation sémantique (Nooralahzadeh *et al.*, 2016). Les propriétés se restreignent à celles d'appartenance "pv:belongsTo" dont l'extraction automatique obtient les meilleurs résultats. L'utilisateur peut cliquer sur une entité dans le texte pour générer un graphe RDF, à la volée. Ce graphe montre toujours l'entité choisie comme nœud central.

– **Interaction avec le graphe.** L'utilisateur peut interagir avec le graphe : 1) en cliquant sur les nœuds, ce qui permet de lancer une nouvelle requête à la volée dans deux sources du web (DBpedia et NetSent) ; 2) en survolant les nœuds pour afficher des images le cas échéant (logos des marques par exemple).

D'un point de vue technique (*cf.* figure 3), le contenu textuel de chaque page web parcouru avec le navigateur Chrome subit un pré-traitement qui consiste

---

4. NetSent est une base de connaissances fournissant des résultats d'analyse d'opinion réalisée sur des forums de cosmétiques, développée en collaboration avec Holmes Semantic Solutions : http://www.ho2s.com/fr /

principalement à éliminer temporairement les balises HTML qui ne sont pas tolérées par l'analyseur morphosyntaxique. Le texte brut en résultant est soumis à l'analyseur morphosyntaxique Holmes. Le résultat de cette analyse est utilisé par le système à base de règles Renco qui a pour objectif de reconnaître les entités nommées. Pour chacune de ces entités, un module de liage permet de tisser les liens avec leurs équivalents (non ambiguë) présents dans DBpedia. Le module d'extraction de relations permet ensuite de construire des triplets RDF. Ces triplets sont générés automatiquement puis insérés dans la base de connaissances temporaire (API Jena) au fur et à mesure de la consultation des pages web. Une validation manuelle des nouveaux triplets est effectuée par l'expert du domaine (en l'occurrence ici un expert du domaine de la cosmétique) afin d'assurer la qualité des données (voir section 4.3). Une interface graphique permet de faciliter la tâche à l'expert. Après validation, les triplets sont reversés dans la base de connaissances finale. Cette dernière est interrogée, en parallèle d'autres bases du web sémantique (DBpedia et Netsent) pour construire le graphe RDF qui est restitué à l'utilisateur sur demande.

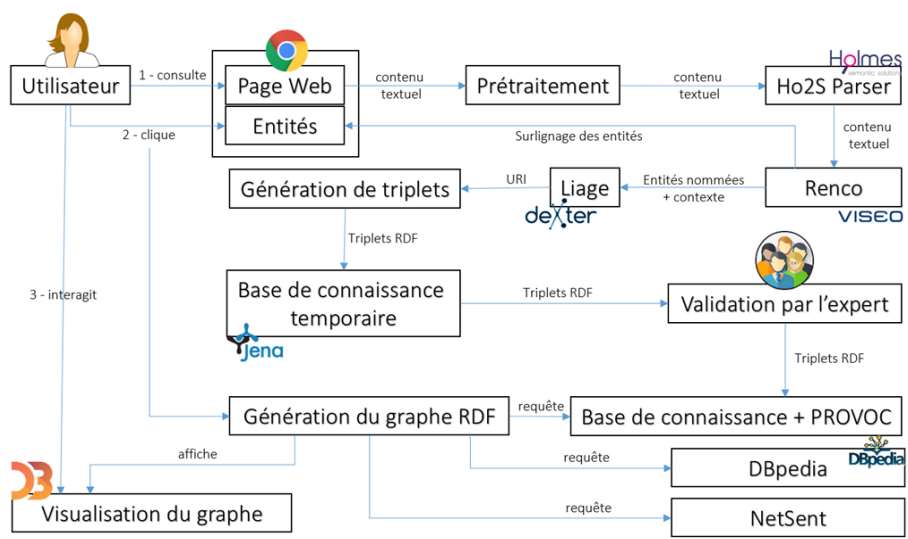

*Figure 3. Chaîne de traitement SMILK*

### 4.2. Exemple de scénario « utilisateur »

L'utilisateur demande au plugin SMILK d'analyser les pages web en vue d'identifier des entités pertinentes du domaine de la cosmétique. Soit « Chanel, haute couture et parfums de luxe à la française » le titre de la page web concernée[5].

---

5. http://www.vogue.fr/communaute/wiki-de-la-mode/articles/chanel-haute-couture-et-parfums-de-luxela-franaise/20583Q211gp7CAxrGaWJD.99

De façon quasi instantanée, l'utilisateur obtient une liste des entités présentes dans le texte qui sont des instances typées par des classes de ProVoc, en particulier : les groupes (ex : Chanel), les marques (ex : Fendi, Prada, Kenzo, ...), les gammes de produits (ex : Chanel maquillage) et les noms de produits (ex : Chanel n°5). Les entités repérées sont surlignées de différentes couleurs et sont synthétisées dans un encadré qui peut être visible ou non selon le souhait de l'utilisateur. L'encadré permet de donner un aperçu du contenu de la page web sans avoir à la parcourir. Un aperçu de cette étape, réalisée en temps réel, est donné en figure 4.

*Figure 4. Un résultat de l'extraction des entités d'intérêt sur une page web de Vogue.fr*

L'utilisateur peut ensuite cliquer sur une entité, par exemple « Chanel ». Le clic a pour effet l'affichage d'un graphe RDF dans une fenêtre aux dimensions réduites. Les connaissances liées à Chanel, provenant des bases DBpedia et NetSent (en bleu) ainsi que les connaissances issues de la base générée automatiquement (en marron) au cours des navigations antérieures (tous les utilisateurs confondus) sont affichables par simple clic ou survol des nœuds concernés (*cf.* figure 5).

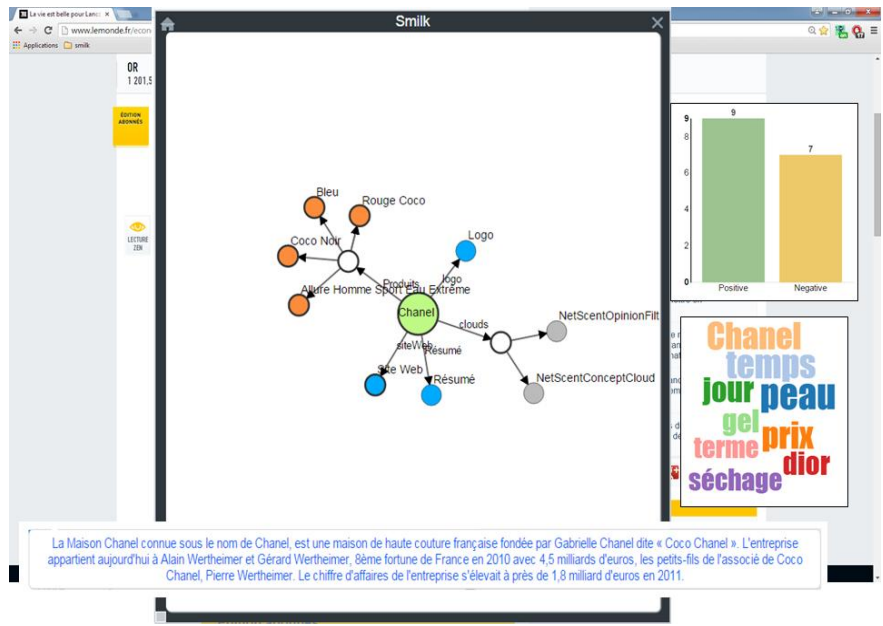

*Figure 5. Liage des entités extraites avec les données de bases de connaissance privées et publiques. En bleu, données provenant de DBpedia ; en gris : données provenant de NetSent ; en orange : données de notre base de connaissances privée*

*4.3. Exemple de scénario « expert »*

Lors de la navigation de l'utilisateur sur le web, des triplets RDF sont générés automatiquement et insérés dans la base de connaissances temporaire. Pour que le système tienne compte de ces données, il requiert une validation des triplets par l'expert du domaine. L'interface restitue le contexte à l'utilisateur afin qu'il soit en mesure de valider ou invalider un triplet (*cf.* figures 6 et 7). L'expert peut accepter le triplet proposé, le modifier, ou le supprimer.

L'expert peut également naviguer dans la base de connaissances en utilisant notre outil SMILK Viewer (*cf.* figure 8). L'accès au graphe de connaissance s'opère en choisissant une entité particulière dans la liste des entités catégorisées par type (groupes, divisions, marques, etc.). L'expert peut ensuite facilement y naviguer et découvrir les connaissances recueillies par le plugin au fil des pages web parcourues, ainsi que des données DBpedia et Netscent s'y rapportant (*cf.* figure 9). Par exemple, en sélectionnant le produit « La Vie est Belle », apparaissent différents composants de cette « Eau de parfum » de « Lancôme » dont du « geraniol » et du « linalool » représentés par l'ontologie ProVoc (individus de la classe « pv:Component »). En cliquant sur « Olivier Polge », le graphe permet de prendre connaissance des produits développés par ce parfumeur. Le base de connaissances est ainsi parcourable dans son intégrité.

*Figure 6. Interface montrant les articles nécessitant la validation de triplets.*

*Figure 7. Interface de validation des triplets par un expert*

*Figure 8. Aperçu des entités nommées dans l'application de visualisation de la base de connaissances*

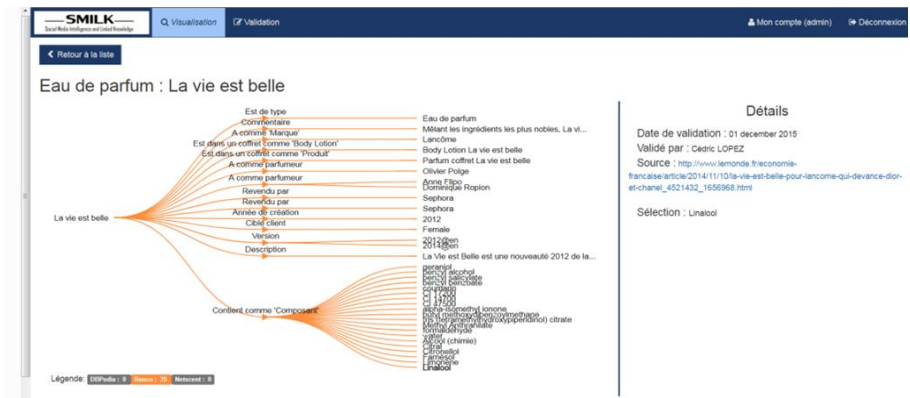

*Figure 9. Visualisation des composants du produit « La Vie est Belle »*

### 5. Conclusion

Le laboratoire commun SMILK a expérimenté différentes approches à l'interface entre TALN et web sémantique. Dans cet article, nous nous sommes focalisés dans un premier temps sur la construction de l'ontologie ProVoc. Cette ontologie est utilisable pour représenter des catalogues de produits, en considérant des classes telles que les gammes de produits, des paquets de produits vendus en tant qu'unité, ou encore les composants de produits, entre autres. ProVoc n'a pas pour ambition de représenter une taxonomie des produits existants. Les produits représentés par ProVoc peuvent être typés (rdf:type) en utilisant d'autres ontologies telles qu'eClassOWL ou UnspscOWL (Hepp, 2005) (Hepp, Bruijn, 2007) en fonction du cas d'application envisagé. Associée à GoodRelations, ProVoc apporte aux sociétés une meilleure visibilité de leurs produits, plus fine, ainsi qu'une plus grande transparence sur les produits commercialisés qui offre à l'utilisateur/consommateur de nouveaux modes de recherches et la possibilité d'exploiter plus de critères pour exprimer des requêtes répondant à ses attentes.

Dans un second temps, nous avons décrit nos expériences visant à peupler une base de connaissances basée sur l'ontologie ProVoc. L'étape d'extraction de relations décrite dans cet article s'appuie sur un traitement linguistique pour appliquer des règles définies manuellement à partir de différents corpus. La prise en compte de lexiques au sein de règles syntaxiques a permis d'obtenir une précision convenable qui permet, en complément d'autres approches d'extractions de relations, de soumettre les triplets générés à un expert lors d'une phase de validation.

Enfin, nous avons présenté notre application qui prend la forme d'un plugin Chrome. Elle est une démonstration de l'utilisabilité des algorithmes retenus dans le cadre du projet SMILK au sein d'une chaîne de traitement prenant en entrée du texte brut et fournissant en sortie à l'utilisateur des connaissances qui n'apparaissaient pas au premier abord.

Plusieurs directions de recherches s'ouvrent sur chaque étape mentionnée. Il existe maintenant des approches très différentes pour l'extraction de connaissances à partir de texte qui ouvrent aussi de nouvelles questions sur les façons efficaces de les combiner. En complément, les retours utilisateurs, même indirects, peuvent aussi permettre d'améliorer les extractions (*e.g.* nouveaux exemples ou contre-exemples) et les connaissances (*e.g.* contributions ou corrections) dans une forme collective de l'apprentissage et du traitement. Enfin, demeure la possibilité d'intégrer plus de sources, de vocabulaires et de référentiels pour enrichir et structurer les connaissances et piloter leur acquisition.

**Bibliographie**

Alec C., Safar B., Reynaud-Delaître C., Sellami Z., Berdugo U. (2014). Peuplement automatique d'ontologie à partir d'un catalogue de produits. In *25$^e$ journées francophones d'ingénierie des connaissances*, IC 2014, Clermont-Ferrand, France, p. 87–98.

Amardeilh F., Laublet P., Minel J.-L. (2005). Annotation documentaire et peuplement d'ontologie à partir d'extractions linguistiques. In *26$^e$ journées francophones d'ingénierie des connaissances*, IC 2005, Grenoble, p. 100–112.

Ashraf J., Cyganiak R., O'Riain S., Hadzic M. (2011). Open ebusiness ontology usage: Investigating community implementation of goodrelations. In *Proceedings of the 20th international world wide web conference*.

Bachimont, B. (2000). Engagement sémantique et engagement ontologique: conception et réalisation d'ontologies en ingénierie des connaissances. Ingénierie des connaissances: évolutions récentes et nouveaux défis, p. 305-323.

Brickley D., Miller L. (2010). *Foaf vocabulary specification 0.98. namespace document 9 august 2010*, Marco Polo Edition, http://xmlns.com/foaf/spec/20100809.html

Ceccarelli D., Lucchese C., Orlando S., Perego R., Trani S. (2013). Dexter: an open source framework for entity linking. In *Proceedings of the sixth international workshop on exploiting semantic annotations in information retrieval*, CIKM'13, 22nd ACM Int. Conf. on Information and Knowledge Management, San Francisco, CA, USA, p. 17-20.

Gerber D., Hellmann S., Buhmann L., Soru T., Usbeck R., Ngomo A.-C. N. (2013). Real-time rdf extraction from unstructured data streams. In *Proceedings of the 12th international semantic web conference*, ISWC 2013, Sydney, NSW, Australia, vol. 8218, p. 135-150. Springer.

Grüninger M., Fox M. S. (1995). Methodology for the design and evaluation of ontologies. In *Proceedings of the Workshop on Basic Ontological Issues in Knowledge Sharing*, Workshop on Basic Ontological Issues in Knowledge Sharing, IJCAI, Québec, Canada.

Hearst M. A. (1992). Automatic acquisition of hyponyms from large text corpora. In *Proceedings of the 14th conference on computational linguistics*, vol. 2, Association for Computational Linguistics, Nantes, France, p. 539–545.

Hepp M. (2005). EclassOWL: A fully-fledged products and services ontology in owl. *Poster Proceedings of ISWC2005*. Galway.

Hepp M. (2008). Goodrelations: An ontology for describing products and services offers on the web. *Knowledge Engineering: Practice and Patterns*, p. 329–346.

Hepp M., Bruijn J. de. (2007). Gentax: A generic methodology for deriving owl and rdfs ontologies from hierarchical classifications, thesauri, and inconsistent taxonomies. In *European semantic web conference*, Innsbruck, Austria, p. 129–144.

Kiryakov A., Popov B., Terziev I., Manov D., Ognyanoff D. (2004). Semantic annotation, indexing, and retrieval. *Web Semantics: Science, Services and Agents on the World Wide web*, vol. 2, no 1, p. 49–79.

Kumar K., Manocha S. (2015). Constructing knowledge graph from unstructured text. *Self*, 3, p. 4.

Lopez C., Nooralahzadeh F., Cabrio E., Segond F., Gandon, F. (2016). ProVoc: une ontologie pour décrire des produits sur le web. In *IC2016: 27$^e$ Journées francophones d'Ingénierie des Connaissances*, Montpellier, France, p. 61-72.

Lopez C., Osmuk M., Popovici D., Nooralahzadeh F., Rabarijaona D., Gandon F., Segond F. (2016). Du TALN au LOD: Extraction d'entités, liage, et visualisation. In *IC2016: 27$^e$ Journees francophones d'Ingenierie des Connaissances, Montpellier, France (demo paper)*.

Lopez C., Cabrio E., Segond F. (2017). Extraction de relations pour le peuplement d'une base de connaissances à partir de tweets. In *17$^e$ journées francophones extraction et gestion des connaissances*, EGC 2017, Grenoble, France, p. 375-380.

Lopez C., Segond F., Hondermarck O., Curtoni P., Dini L. (2014). Generating a resource for products and brandnames recognition. Application to the cosmetic domain. In *Proceedings of the ninth international conference on language resources and evaluation*, LREC 2014, Reykjavik, Iceland, p. 2559–2564. European Language Resources Association (ELRA).

McDonald D. (1996). Internal and external evidence in the identification and semantic categorization of proper names. *Corpus processing for lexical acquisition*, p. 21–39.

Nebhi K. (2013). A rule-based relation extraction system using DBpedia and syntactic parsing. *Proceedings of the NLP & dbpedia workshop co-located with the 12th int. semantic web conference (ISWC 2013)*, Sydney, Australia, vol. 1064, p. 74-79.

Nooralahzadeh F., Lopez C., Cabrio E., Gandon F., Segond F. (2016). Adapting semantic spreading activation to entity linking in text. In *Proceedings of the 21st International Conference on Applications of Natural Language to Information Systems*, NLDB 2016, Salford, UK, vol. 9612, p. 74–90. Springer.

Tounsi M., Lopez C., Faron Zucker C., Cabrio E., Gandon F., Segond F. (2017). Peuplement d'une base de connaissances par annotation automatique de textes relatifs à la cosmétique. In *28$^e$ Journées francophones d'Ingénierie des Connaissances,* IC 2017, p. 104-114. Caen, France.

Urieli A. (2013). *Robust french syntax analysis: reconciling statistical methods and linguistic knowledge in the Talismane toolkit*. Thèse de doctorat non publiée, Université Toulouse le Mirail-Toulouse II.

Uschold M., Gruninger M. (1996). Ontologies: Principles, methods and applications. *The knowledge engineering review*, vol. 11, no 2, p. 93-136.